\documentclass[12pt]{article}

\usepackage{latexsym}
\usepackage{amsthm}
\usepackage{amsmath,amssymb,amsfonts,bm,amsbsy,bbm}
\usepackage{epsfig}
\usepackage{graphicx,rotating,lscape}
\usepackage{verbatim}
\usepackage{geometry}
\usepackage{setspace}
\usepackage[dvipsnames]{xcolor}
\usepackage{float}
\usepackage{physics}

\usepackage[
    colorlinks=true,
    linkcolor=blue!70!black,    
    citecolor=Sepia,    
    urlcolor=teal!80!black      
]{hyperref}

\usepackage{booktabs}

\usepackage{algorithm}
\usepackage{algpseudocode}

\usepackage{multirow}
\usepackage{longtable}
\usepackage{natbib}

\usepackage{enumitem}
\makeatletter
\newcommand{\thickhline}{%
  \noalign{\global\let\saved@LT@hlines\LT@hlines}%
  \multispan{\LT@cols}{\leaders\hrule height 1pt\hfill}\cr
}
\makeatother

\usepackage{ulem}

\def\bse{\begin{eqnarray*}}
	\def\ese{\end{eqnarray*}}
\def\be{\begin{eqnarray}}
	\def\ee{\end{eqnarray}}
\def\bsq{\begin{equation*}}
	\def\esq{\end{equation*}}
\def\bq{\begin{equation}}
	\def\eq{\end{equation}}
\def\bi{\begin{itemize}}
	\def\ei{\end{itemize}}

\def\sumi{\sum_{i=1}^n}

\def\vech{{\rm vech}}
\def\idxset{{\cal I}}

\def\ol{\overline}
\def\wh{\widehat}

\def\pr{\hbox{pr}}

\def\calB{{\cal B}}

\def\trans{^{\rm T}}

\def\n{\nonumber}

\newcommand\indep{\protect\mathpalette{\protect\independenT}{\perp}}
\def\independenT#1#2{\mathrel{\rlap{$#1#2$}\mkern2mu{#1#2}}}

\def\bb{{\boldsymbol\beta}}

\def\bSigma{{\boldsymbol\Sigma}}

\def\0{{\bf 0}}
\def\1{{\bf 1}}
\def\X{{\bf X}}
\def\x{{\bf x}}
\def\Z{{\bf Z}}
\def\z{{\bf z}}

\def\W{{\bf W}}
\def\w{{\bf w}}
\def\U{{\bf U}}

\def\V{{\bf V}}
\def\v{{\bf v}}
\def\S{{\bf S}}
\def\s{{\bf s}}

\def\B{{\bf B}}

\def\b{{\bf b}}
\def\A{{\bf A}}
\def\a{{\bf a}}

\def\g{{\bf g}}

\def\I{{\bf I}}
\def\M{{\bf M}}
\def\m{{\bf m}}

\def\Y{{\bf Y}}

\def\R{{\bf R}}

\def\t{{\bf t}}

\def\boxit#1{\vbox{\hrule\hbox{\vrule\kern6pt\vbox{\kern6pt#1\kern6pt}\kern6pt\vrule}\hrule}}

\def\MVN{{\cal N}}

\def\invT{^{-\rm T}}

\def\var{\hbox{var}}

\newtheorem{theorem}{Theorem}[section]

\allowdisplaybreaks

\begin{document}
\title{Quantile regression with measurement errors}

\author{
    Mushan Li\textsuperscript{1}, Yanyuan Ma\textsuperscript{1,*}  and  Liqun Wang\textsuperscript{2} \\[0.3cm]
    \small \textsuperscript{1}Department of Statistics, Pennsylvania State University, University Park, PA 16802, USA \\
    \small E-mail: \texttt{mzl5849@psu.edu} (M. Li), \texttt{yzm63@psu.edu} (Y. Ma) \\[0.2cm]
    \small \textsuperscript{2}Department of Statistics, University of Manitoba, Winnipeg, Canada \\
    \small E-mail: \texttt{liqun.wang@umanitoba.ca}
}

\date{}

\maketitle
\renewcommand{\thefootnote}{\fnsymbol{footnote}}
\footnotetext[1]{Corresponding author.}
\renewcommand{\thefootnote}{\arabic{footnote}}
\setcounter{footnote}{0}

\thispagestyle{empty}

\begin{abstract}
We devise a novel estimator for a general quantile regression model with normal measurement errors in the covariates. The method is applicable to both linear and nonlinear quantile regressions and does not impose the quantile requirement on multiple quantile levels simultaneously. 
We circumvent the difficulties caused by discontinuity in quantile regression through kernel smoothing, and overcome the nonlinearity inherent in quantile regression via considering extension to the complex domain  and moment generating functions. 
We show that the resulting estimator achieves the standard root-$n$ consistency and asymptotic normality under mild conditions. The performance of the proposed method is illustrated via numerical simulations and a real data example related to Cherry Blossom times in Japan  in 2024. 
This is the first consistent estimator in a general quantile regression problem with normal measurement errors.
\end{abstract}



\section{Introduction}
In studying the relation between a response variable and the
covariates, mean and quantile  regressions are the most often considered
forms of analysis. Compared to mean regression, quantile regression has its
advantages in that it can provide a more global picture if multiple
quantiles are considered and is more robust, hence has received much
attention \citep{koenker1978regression,  koenker2017quantile}. Indeed, quantile regression
has been widely used in many fields, such as  bioinformatics, education, 
finance and economics, and medical and health sciences \citep{yu2003quantile,
koenker2017handbook, wang2018wild, li2019map, mawei2012, de2019adapted}.
While linear quantile regression models have been extensively studied
in the literature,  
research on general nonlinear quantile regression models is relatively
rare. 

In practice, often some predictors are not directly observable or are
measured with substantial errors. It is well-known that simple substitution
of the surrogate data for the latent variables will result in
attenuated and inconsistent estimators in quantile as well as mean
regression models \citep{fuller2009measurement,
  carroll2006,wei2009quantile}.  
Interestingly, although the measurement error problem in mean regression
has been well studied and to a large extend fully solved
\citep{carroll2006, garciama2017, li2019semiparametric,
  li2024robust, li2024update}, the situation about quantile regression is much more
bleak due to the difficult nature of the problem. One of the main
challenges is that,  unlike the expectation operator, the quantile
function is nonlinear and non-smooth so that the true covariates and measurement
errors cannot be separated. 

Several authors have tackled the measurement error problem in quantile
regression setup, however, so far the existing works are limited to
the linear models only.
\cite{he2000quantile} studied the orthogonal distance quantile regression in a linear
measurement error model. Their method relies on the assumption that
the measurement and regression errors have a joint spherically
symmetric distribution. This implies that the model error and
measurement error follow the same distribution, ensuring
identifiability for regression coefficients. This method is further
fused with other approaches to handle more complex problems
\citep{mayin2011, wumayin2015}.  
Alternatively, \cite{wei2009quantile} proposed a deconvolution-based
method that requires all conditional
quantiles to satisfy the linear relation, even though only regression
at a single quantile is of interest. It results in much more stringent
model assumption and heavy numerical computations.
Similarly, \cite{firpo2017measurement}
also used a deconvolution procedure to recover the
conditional density of the error-prone covariate and then construct
estimating equations. 
Although the method only focuses on one quantile level, it still
engages heavy computation due to the deconvolution step involved.
A third type of approach is given by \cite{wang2012corrected}, where they proposed a
corrected-loss estimation (CLE) for linear quantile regression which
only requires estimation at the quantile of interest, however, it is
restricted to univariate mismeasured covariate.
Further, \cite{guan2017instrumental} and \cite{yang2020composite} used
instrumental variable method to estimate linear quantile regression
models.  Apart from the limitation to linear models under various
restrictive assumptions, many of the published works so far only
provide the estimates for the slope parameters and not for the
intercept parameter in the linear model.
More importantly, as far as we are aware, no work exists on nonlinear
quantile regression, with measurement errors in either single or
multiple covariates. 
This gap is not surprising. Linear quantile regression with
measurement error is already difficult because the measurement error
complicates the estimation process.   Nonlinear quantile regression,
without measurement error, is also difficult due to the complex forms
of the associated estimating functions.  Combining both
aspects---nonlinear quantile regression and measurement
error---doubles the difficulty of the problem.  
This work aims to address this challenge by developing a method that
handles measurement errors in general quantile regression, with linear
quantile regression as a special case.

We consider a general possibly nonlinear quantile regression model
where one or more covariates are subject  to normal measurement
error. We propose an unbiased estimating equation method by introducing
random variables in the complex number domain to cancel out the
measurement errors in the 
covariates. Although this method has been used in mean regression with
smooth regression functions  \citep{stefanski1989}
and is closely linked to the idea of simulation-extrapolation \citep{stefanski1995simulation},
 its use in quantile regression setup is not straightforward or immediately obvious. 
To circumvent the difficulties caused by the discontinuity
in quantile regression, which breaks the pivotal assumption needed for
the methodological development,  we further adopt
a kernel approximation  technique. Our end result is a computationally
efficient estimation method that accounts
for measurement errors in general linear and  nonlinear quantile regression models.

\section{Methods}\label{sec:method}
\subsection{Model and Method}\label{sec:21}
Consider a general quantile regression model
\be \label{themodel}
Y=m(\X,\Z,\bb)+\epsilon,
\ee
where $m(\X, \Z, \bb)$ is  the $\tau$th conditional  quantile
  function of $Y$ given $\X,\Z$,  $\X$ and $\Z$ are
vectors of the covariates measured with and without error
respectively, $\bb$ is the vector of unknown parameters, and 
$Q_{\epsilon\mid\X,\Z}(\tau,\X,\Z)=0$ with $\tau\in (0,1)$ being a 
prespecified quantile level of interest.
Here, $Q_{\epsilon\mid\X,\Z}(\tau,\X,\Z)\equiv \inf\{ r \in \R: \pr(\epsilon\le
  r\mid \X,\Z)\ge\tau\}$ is the quantile function of $\epsilon$ given
  $\X, \Z$. Because $\epsilon\equiv Y-m(\X,\Z,\bb)$,
  $\bb$ and subsequently $\epsilon$ depend on $\tau$.
Following the measurement error literature
\citep{fuller2009measurement, carroll2006,wei2009quantile,
  wang2012corrected}, we 
also assume a classical measurement error model
$\W=\X+\bSigma^{1/2}\U$, where 
$\U\indep(\X, \Z, \epsilon)$ and $\U\sim \MVN(\0, \I)$. Here,
$\bSigma$ is either known or  estimable and $\indep$ denotes
statistical independence. 
We observe iid data $(\W_i,
\Z_i, Y_i), i=1, \dots, n$ 
and  aim at estimating $\bb$.
In model \eqref{themodel}, 
we assume that at any $\Z,\bb$, the complex extension
of $m(\X, \Z, \bb)$ and its derivative with respect to $\bb$, i.e., 
$\m'_\bb(\X, \Z, \bb)$, are
 analytic on the entire complex domain.

Let $\rho_\tau(t)$ be the check function, i.e.,
$\rho_\tau(t)\equiv I(t\ge0)\tau t
- I(t<0)(1-\tau)t$. We define the corresponding ``derivative''
function 
$g_\tau(t)\equiv I(t\ge 0)\tau 
- I(t<0)(1-\tau)+t\delta(t)$, where $\delta(\cdot)$ is the Dirac delta
function.  Because $g_\tau(t)$ is a discontinuous function of $t$,
we  consider a smoothed version of $g_\tau(t)$, defined as
$\psi(t)\equiv\int g_\tau(s)k_h(t-s)ds +tk_h(t)
=\tau-1+K(t/h)+tk_h(t)$. Here, $k_h(t)\equiv k(t/h)/h$ and
$k(\cdot)$ is a symmetric kernel function with its
  complex extension analytic on the entire complex domain, 
  and $K(\cdot)$ is the corresponding cdf.
Examples of a qualified kernel function includes
    the Gaussian pdf $k(t)=e^{-t^2/2}/\sqrt{2\pi}$
    and  the pdf of the form $k(t)=e^{-t^{2j}}/j/\Gamma\{1/(2j)\}$,
    where $j$ is a positive integer and $\Gamma(\cdot)$ is the Gamma
    function. In practice, we can simply 
    use the  Gaussian kernel for convenience.
Note that $\psi(\cdot)$ is also an analytic function on the entire
  complex domain.

If $\X_i$ were available, we could use
$\psi\{Y_i-m(\X_i,\Z_i,\bb)\}\m_\bb'(\X_i,\Z_i,\bb)$ to form
estimating equation for the 
purpose of estimating $\bb$. Without $\X_i$, our goal is to
find a function, say $\S(Y_i, \W_i, \Z_i, \bb, \bSigma)$ so that
$E\{\S(Y_i, \W_i, \Z_i, \bb, \bSigma)\mid Y_i,\X_i, \Z_i\}=
\psi\{Y_i-m(\X_i,\Z_i,\bb)\}\m_\bb'(\X_i,\Z_i,\bb)$, hence
we can still form an estimating equation for $\bb$ by using
$\S(Y_i, \W_i, \Z_i, \bb, \bSigma)$. To this end, we adopt the
idea of \cite{stefanski1989}  to ``cancel'' the measurement
error $\U_i$ in $\X_i+\U_i$ by adding a ``cancel variate'' $\sqrt{-1}\V_i$. 
More formally, for notational brevity, define
  \bse
\textstyle 
  \S(Y_i, \W_i, \Z_i, \bb, \bSigma)\equiv
 E[
\psi\{Y_i-m(\W_i+\sqrt{-1}\bSigma^{1/2}\V_i,\Z_i,\bb)\}
\m_\bb'(\W_i+\sqrt{-1}\bSigma^{1/2}\V_i,\Z_i,\bb)
\mid
\W_i,\Z_i,Y_i],
\ese
where $\V_i \sim \MVN(\0,\I)$ is a random vector independent of all
other variables.
We propose an estimator for $\bb$ under the measurement error context
by solving
\be\label{eq:esteqc}
\sumi \S(Y_i, \W_i, \Z_i, \bb, \bSigma)
=\0.
\ee
In practice, when the
expectation is difficult to compute, we can replace it by sample
average, i.e.,
we use
$B^{-1}\sum_{b=1}^B
\psi\{Y_i-m(\W_i+\sqrt{-1}\bSigma^{1/2}\V_{ib},\Z_i,\bb)\}
\m_\bb'(\W_i+\sqrt{-1}\bSigma^{1/2}\V_{ib},\Z_i,\bb)$, where $\V_{ib},
b = 1,2,...,B$ is an independent sample from $\MVN(\0,\I)$, to replace
the expectation above for a very large $B$, 
and retain only the real part to form the estimating equation.

A key feature of our method is that it bypasses 
  estimating the distribution of the error-prone variable $\X_i$,
  which involves deconvolution and is known to have very slow convergence
  rate \citep{carrollhall1988}. Avoiding estimating the distribution
  of $\X_i$ is a very important advantage
  compared to deconvolution based methods such as
  \cite{wei2009quantile} and \cite{firpo2017measurement}.

\subsection{Theory}\label{sec:theory}
Let $\X\in\R^d$ and $\s$ be a length $d$ vector of nonnegative integers.
We define $\s!\equiv
\prod_{l=1}^d s_l!, |\s|\equiv\sum_{l=1}^d s_l$.
For any function $\g(\v): \R^d\to \R^d$, we define $\partial
^\s\g(\v)/\partial \v^s\equiv \partial^{|\s|}\g(\v)/\prod_{l=1}^d\partial
v_l^{s_l}$
and $\v^\s\equiv(v_1^{s_1}, \dots, v_d^{s_d})\trans$. For any two
length $d$ vectors $\a,\b$, let $\a*\b\equiv \prod_{l=1}^da_lb_l$.

For preparation, note that if a function $\g(\cdot)$ is analytic
  on the entire complex domain, it is
infinitely differentiable everywhere. Thus,
we have
\be\label{eq:len}
&&E\{\g(\W+\sqrt{-1}\bSigma^{1/2}\V,\Z,Y)\mid\X,\Z,Y\}\n\\
&=&E\{\g(\X+\bSigma^{1/2}\U+\sqrt{-1}\bSigma^{1/2}\V,\Z,Y)\mid\X,\Z,Y\}\n\\
&=&\g(\X,\Z,Y)+\sum_{k=1}^\infty
\sum_{|\s|=k}
\frac{\partial^{k}\g(\X,\Z,Y)}{\s!\partial\X^{\s}
}E[
\{\bSigma^{1/2}(\U+\sqrt{-1}\V)\}^\s]\n\\
&=&\g(\X,\Z,Y),
\ee
where the last equality holds because the moment generating function
\bse
M_{\bSigma^{1/2}(\U+\sqrt{-1}\V)}(\t)&=&
E [\exp\{\t\trans\bSigma^{1/2}(\U+\sqrt{-1}\V)\}]\\
&=&1+\sum_{k=1}^\infty
\frac{1}{k!}E[
\{\t\trans\bSigma^{1/2}(\U+\sqrt{-1}\V)\}^k]\\
&=&1+\sum_{k=1}^\infty\sum_{|\s|=k}
\frac{1}{\s!}\t^{\s}* E[
\{\bSigma^{1/2}(\U+\sqrt{-1}\V)\}^\s]\\
&=&M_{\U}(\bSigma^{1/2}\t) M_{\sqrt{-1}\V}(\bSigma^{1/2}\t)\\
&=&\exp(\t\trans\bSigma\t)\exp(-\t\trans\bSigma\t)\\
&=&1,
\ese
which implies $E[
\{\bSigma^{1/2}(\U+\sqrt{-1}\V)\}^\s] =\0$ for all
$|\s|>0$. The calculation result in \eqref{eq:len} establishes
  that adding the random component $\sqrt{-1}\V$ to $\U$ cancels the
measurement error  effect caused by $\U$, i.e., it is a ``cancel variate''.

Now consider the estimating function $\S(\cdot)$ defined above \eqref{eq:esteqc}.
It follows that
\bse
&&E\{\S(Y_i, \W_i, \Z_i, \bb, \bSigma)\}\n\\
&=&E[
\psi\{Y_i-m(\W_i+\sqrt{-1}\bSigma^{1/2}\V_i,\Z_i,\bb)\}
\m_\bb'(\W_i+\sqrt{-1}\bSigma^{1/2}\V_i,\Z_i,\bb)]\n\\
&=&E(E[
\psi\{Y_i-m(\W_i+\sqrt{-1}\bSigma^{1/2}\V_i,\Z_i,\bb)\}
\m_\bb'(\W_i+\sqrt{-1}\bSigma^{1/2}\V_i,\Z_i,\bb)\mid \X_i, \Z_i,
Y_i])\n\\
&=&E[
\psi\{Y_i-m(\X_i,\Z_i,\bb)\}
\m_\bb'(\X_i,\Z_i,\bb)]\n\\
&=&\textstyle 
E[\{\tau-1+E(K[\{Y_i-m(\X_i,\Z_i,\bb)\}/h]
+\{Y_i-m(\X_i,\Z_i,\bb)\} k_h\{Y_i-m(\X_i,\Z_i,\bb)\}
\mid \X_i, \Z_i)\}\n\\
&&\times\m_\bb'(\X_i,\Z_i,\bb)],
\ese
where the first equality uses the definition of
$\S(Y_i, \W_i, \Z_i, \bb, \bSigma)$,
the third equality is because of \eqref{eq:len},
and the fourth equality uses the definition of $\psi(\cdot)$.
Write $C_2\equiv\int t^2k(t)dt/2$, and let
$f'_{\epsilon\mid\X,\Z}(\epsilon,\X,\Z)$  and
$f''_{\epsilon\mid\X,\Z}(\epsilon,\X,\Z)$ respectively denote the
first and second derivatives of 
$f_{\epsilon\mid\X,\Z}(\epsilon,\X,\Z)$ with respect to $\epsilon$.
With a simple change of variable calculation and a Taylor
expansion, we continue the above calculation to obtain
\be\label{eq:h2c}
&&E\{\S(Y_i, \W_i, \Z_i, \bb, \bSigma)\}\n\\
&=&
E[\{\tau-1+ \int_{-\infty}^\infty K(t) f_{\epsilon\mid\X,\Z}(ht, \X_i, \Z_i) hdt\}
\m_\bb'(\X_i,\Z_i,\bb)] +O(h^2)\n\\
&=&\textstyle 
E[\{\tau-1+ K(t) F_{\epsilon\mid\X,\Z}(ht, \X_i, \Z_i)\bigg\rvert_{-\infty}^\infty
 - \int_{-\infty}^\infty k(t)F_{\epsilon\mid\X,\Z}(ht, \X_i, \Z_i)dt \} 
 \m_\bb'(\X_i,\Z_i,\bb)] +O(h^2)\n\\
&=&E([\tau -  \int_{-\infty}^\infty k(t) \{F_{\epsilon\mid\X,\Z}(0, \X_i, \Z_i) + f_{\epsilon\mid\X,\Z}(0, \X_i, \Z_i)ht \}dt]\m_\bb'(\X_i,\Z_i,\bb)) +O(h^2)\n\\
&=&O(h^2),
\ee
where the second equality performs integration by parts, the
third equality follows from the definition of $\tau$, and 
the last equality is due to the symmetry
property of the kernel function $k(\cdot)$.

 To derive the theoretical properties of our proposed estimator, we
 assume the following regularity conditions.

\begin{enumerate}[label=(C\arabic*),ref=(C\arabic*),start=1]
\item \label{con:uniroot}
$E\{\S(Y,\W,\Z, \bb, \bSigma)\}=\0$ has a unique solution in a
neighborhood of the true parameter $\bb$.
\item  \label{con:compact}
  The parameter space of $\bb$, $\calB$, is a compact set.
\item  \label{con:bddregfunc}
  The regression function  $m(\x,\z,\bb)$  has second derivative with
  respect to $\bb$, the matrix \\
  $E\{f_{\epsilon\mid\X,\Z}(0,\X,\Z)
  \m_\bb'(\X,\Z,\bb)^{\otimes2}\}$ is nonsingular, and
  $E\{\|\m_\bb'(\X,\Z,\bb)\|^2\}$ is bounded.
\item \label{con:diferrfunc}
  The conditional density function $f_{\epsilon\mid\X,\Z}$ is twice
  differentiable with respect to $\epsilon$ at $\epsilon=0$.
\item  \label{con:ker}
  The kernel function $k(\cdot)$ is positive and symmetric when evaluated on
    the real line, and  analytic on the entire complex domain.
\item  \label{con:nh}
  The bandwidth $h$ satisfies $nh^4\to0, nh\to\infty$.
\end{enumerate}
Conditions \ref{con:uniroot} and \ref{con:compact} are standard
conditions in establishing consistency and are routinely
assumed. Conditions \ref{con:bddregfunc}  and \ref{con:diferrfunc} require some
smoothness, boundedness and nonsingularity, which are also standard
requirements to exclude pathological situations. Conditions
\ref{con:ker} and \ref{con:nh} are about the kernel function and
bandwidth, both are within our control hence can be satisfied by
choice. Specifically, the  analytic property of the kernel function is
required due to the need of taking infinitely many derivatives in
analyzing the moment generating function  on the complex domain
  and maintain the equality relation,
while the upper and
  lower bounds of how fast the bandwidth goes to zero correspond
  respectively to the need of controlling the bias and variance of the
final estimator. All these conditions are very mild conditions and are
usually assumed in semiparametric models.
Here, although seemingly we are dealing with a parametric quantile function
$m(\X,\Z,\bb)$, the quantile regression with measurement error is 
inherently a semiparametric model hence
these conditions are also required.

\begin{theorem}\label{th:classic}
Let $\wh\bb_s$ solve the estimating equation \eqref{eq:esteqc}.  Then
under Conditions\ref{con:uniroot}-\ref{con:nh}, $\wh\bb_s$ is a consistent estimator of $\bb$.
Further $\sqrt{n}(\wh\bb_s-\bb) \to \MVN(\0, \A^{-1}\B_{1}\A\invT)$ in distribution when
$n\to\infty$, where 
$\A=E\left\{f_{\epsilon\mid\X,\Z}(0,\X,\Z)
\m_\bb'(\X,\Z,\bb)^{\otimes2}\right\}$ 
and
$\B_{1}=
E[\{
E([
\tau-1+I\{Y-m(\W+\sqrt{-1}\bSigma^{1/2}\V,\Z,\bb)\ge0\}]
\m_\bb'(\W+\sqrt{-1}\bSigma^{1/2}\V,\Z,\bb)\mid
\W,\Z,Y)\}^{\otimes2}]$.
\end{theorem}
The proof of Theorem \ref{th:classic}  is given in Appendix \ref{apdx:th1pf}.

\subsection{When \texorpdfstring{${\bSigma}$}{} is unknown}
For  simplicity in the above derivation, we have assumed $\bSigma$ to be
  known, while in practice, $\bSigma$ is often unknown and needs to be
  estimated. We now return to the case with
  replicate measurements for $\X$. 
  The general idea is to obtain new measurements and
$\wh\bSigma$ based on the replicates, and then  substitute them for
$\W, \bSigma$ in our methods 
described above. We now describe this step in detail.

Assume for each subject $i \in \{1, ..., n\}$, there are $m(\ge 2)$
replicates $W_{ij}, j=1, \dots, m$,  such that
$\W_{ij} = \X_i + m^{1/2}\bSigma^{1/2}\U_{ij}$, where
$\U_{ij}\indep(\X_i, \Z_i, \epsilon_i)$, $\U_{ij}\indep\U_{ik}$ for
$j\ne k$ and $\U_{ij}\sim \MVN(0, 
\I)$. 
Let $\ol\W_i \equiv m^{-1}\sum_{j=1}^m\W_{ij}$.
Then, $\ol\W_i=\X_i+\bSigma^{1/2}\U_i$, where
$\U_i= m^{-1/2}
\sum_{j=1}^m\U_{ij}\sim \MVN(0,\I)$.
Further, let $\M_i \equiv
(m-1)^{-1}\sum_{j=1}^{m} (\W_{ij} - \ol \W_i)^{\otimes2}$ and we
estimate $\bSigma$ by $\wh\bSigma\equiv(mn)^{-1}\sumi \M_i$.  Then
\be \label{eq:sigc}
\sqrt{n}(\wh\bSigma - \bSigma) 
=\sqrt{n}(\frac{1}{nm}\sumi \M_i - \bSigma ) 
=  \frac{1}{\sqrt{n}}\sumi (m^{-1}\M_i - \bSigma).
\ee

\begin{theorem}\label{th:classichat}
Let $\wh \bSigma$ be defined as above. Replace
$\W_i, \bSigma$ by
 $\overline\W_i, \wh\bSigma$ 
 in \eqref{eq:esteqc} and let
$\wh\bb$ be its root.
Under  Conditions \ref{con:uniroot}-\ref{con:nh}, $\wh\bb\to\bb$ in probability and
$\sqrt{n}(\wh\bb-\bb) \to\MVN\{\0, \A^{-1}(\B_{1}+\B_{2})\A^{-\rm
  T} \}$  in distribution, when $n\to\infty$, where  
$\A, \B_{1}$ are 
identical to those  in Theorem \ref{th:classic}, and
\be\label{eq:Bsig}
\B_{2} = E\left\{\frac{\partial\,\S(\Y_i, \ol\W_i, \Z_i, \bb,
      \bSigma)}{\partial \vech(\bSigma)\trans}\right\}E\left\{\vech (m^{-1}\M_i - \bSigma)^{\otimes2}\right\}E\left\{\frac{\partial\,\S(\Y_i, \ol\W_i, \Z_i, \bb,
      \bSigma)\trans}{\partial \vech(\bSigma)}\right\}.
\ee
\end{theorem}
It is easy to see that the extra variation of $\B_2$ in
(\ref{eq:Bsig}) is due to the estimation of $\bSigma$. 
The proof of Theorem \ref{th:classichat} is given in Appendix \ref{apdx:th2pf}.

\section{Simulation Studies}\label{sec:simu}
We conducted comprehensive simulation studies to evaluate the
finite sample performance of the proposed method in
various scenarios.
 Python code for 
  all simulation studies and the data
  application is publicly available on GitHub at
  \url{https://github.com/carayms/QRiV}.
In each scenario,  we consider five
quantile levels $\tau=0.05, 0.25, 0.5, 0.75$ and 0.95
respectively.  We considered both linear and nonlinear
quantile regression models, in combination with both independent and 
dependent model errors.  In each simulation 1000 replications were
carried out.

\subsection*{Simulation 1}
In the first example, we considered a  linear quantile
  regression model
$Y = \beta_0 + \beta_1 X + \epsilon$ with
$(\beta_0, \beta_1) = (1, 1)$,
where $X$ was generated from a uniform distribution on
 the interval $(5, 5+2\sqrt{3})$.
The model error $\epsilon$ was 
generated from a  normal distribution with standard deviation 0.5 and
$\tau$th quantile 0, and
the measurement error is also normal with standard deviation 0.5.
This simulation setting is similar to \cite{wang2012corrected},
and we generated the data  with sample size $n=200$. 

We implemented the proposed method using the Gaussian
kernel function, i.e., $k(\epsilon)=
\exp(-\epsilon^2/2)/\sqrt{2\pi}$, hence 
$K(\epsilon) = \Phi(\epsilon)$. We set the bandwidth
$h=c\wh \sigma_r n^{-1/3}$, where $\wh \sigma_r$ is the estimated
standard
deviation of the residuals from the naive quantile regression fit. We
set $c=6.5$ for quantiles 
$0.25, 0.5, 0.75$ and $c=7.5$ for the more extreme quantiles
$0.05, 0.95$. 
In calculating expectation with respect to $\V$ in the construction of
the estimating equation, we used Gauss–Hermite quadrature method  to
approximate the integrals.
As comparisons,  we also implemented the naive method in which we
ignored the measurement error and performed standard quantile
regression. We also implemented the method proposed in
\cite{wei2009quantile}, which handles linear quantile regression with
measurement errors, but assuming the linear quantile relation holds for the
all quantile levels between 0 to 1. We also implemented  the
corrected-loss estimation (CLE)  of \cite{wang2012corrected}, 
which is designed only for linear quantile regression. In implementing
CLE, we 
used the same bandwidths as in our proposed method.  In all methods, 
the standard deviation is estimated based on 100 bootstrap samples,
which is used to calculate the 95\% confidence intervals subsequently.

We present the results of the four methods in
Table \ref{tab:1indep}.  These results show that all three
measurement-error-correction methods perform 
well in terms of bias and coverage probabilities at all quantile
levels. Moreover, the simulation and estimated standard deviation by
the asymptotic formula of our proposed estimator are very
close. Overall, the CLE has slightly lower standard deviation and the
coverage rate of Wei-Carroll's method is slightly lower than the nominal
level of $95\%$. As expected, the naive method led to serious bias on estimation
and the coverage probability is severely distorted.

\begin{longtable}{cccccccccc}
\caption{Results of Example 1 (Independent). ``bias'', ``std'', 
  ``$\wh{\rm std}$'', and ``cvg'' are respectively the bias, sample
  standard deviation, estimated standard deviation, the coverage of
  the 95\% confidence intervals of the estimators
  based on 1000 repetitions, and sample size $n=200$.} \label{tab:1indep} \\

\thickhline
Method & \multicolumn{2}{c}{Proposed }&  \multicolumn{2}{c}{Naive }&  \multicolumn{2}{c}{Wei-Carroll's } & \multicolumn{2}{c}{CLE }\\
       & $\beta_0$ & $\beta_1$ & $\beta_0$ & $\beta_1$
       & $\beta_0$ & $\beta_1$& $\beta_0$ & $\beta_1$\\
\thickhline
\endfirsthead

\multicolumn{10}{c}{{\bfseries \tablename\ \thetable{} -- Continued from previous page}} \\[0.5ex]
\thickhline
Method & \multicolumn{2}{c}{Proposed }&  \multicolumn{2}{c}{Naive }&  \multicolumn{2}{c}{Wei-Carroll's } & \multicolumn{2}{c}{CLE }\\
       & $\beta_0$ & $\beta_1$ & $\beta_0$ & $\beta_1$
       & $\beta_0$ & $\beta_1$& $\beta_0$ & $\beta_1$\\
\thickhline
\endhead

\hline
\multicolumn{10}{r}{{Continued on next page...}} \\
\endfoot

\thickhline
\endlastfoot  

$\tau=0.05$ &&&&&&\\ 
bias & -0.0167 & 0.0104 & 1.1129 & -0.2064 & -0.0348 & 0.0114 & 0.0682 & 0.0029 \\ 
std & 0.5963 & 0.0889 & 0.6475 & 0.0958 & 0.6958 & 0.1054 & 0.4221 & 0.0625 \\ 
$\wh{\rm std}$ & 0.6275 & 0.0942 & 0.6554 & 0.0967 & 0.7139 & 0.1069 & 0.4313 & 0.0645 \\ 
cvg & 94.6\% & 95.5\% & 56.0\% & 42.4\% & 93.9\% & 93.8\% & 95.1\% & 95.6\% \\ 
\hline
$\tau=0.25$ &&&&&&\\ 
bias & -0.0695 & 0.0084 & 1.2279 & -0.1997 & 0.124 & -0.0208 & -0.0232 & 0.0021 \\ 
std & 0.5232 & 0.0774 & 0.3731 & 0.0545 & 0.4833 & 0.0725 & 0.4079 & 0.0602 \\ 
$\wh{\rm std}$ & 0.5468 & 0.0815 & 0.3944 & 0.0580 & 0.4848 & 0.0729 & 0.4149 & 0.0616 \\ 
cvg & 95.3\% & 95.3\% & 13.0\% & 8.8\% & 93.8\% & 94.0\% & 95.3\% & 95.4\% \\ 
\hline
$\tau=0.5$ &&&&&&\\ 
bias & -0.0563 & 0.0080 & 1.3330 & -0.1983 & 0.099 & -0.0144 & -0.0177 & 0.0024 \\ 
std & 0.5077 & 0.0746 & 0.3435 & 0.0498 & 0.4661 & 0.0684 & 0.4073 & 0.0598 \\ 
$\wh{\rm std}$ & 0.5212 & 0.0770 & 0.3607 & 0.0530 & 0.4750 & 0.0699 & 0.4122 & 0.0608 \\ 
cvg & 94.7\% & 95.0\% & 5.4\% & 4.6\% & 93.0\% & 92.9\% & 95.4\% & 95.6\% \\ 
\hline
$\tau=0.75$ &&&&&&\\
bias & -0.0481 & 0.0085 & 1.4638 & -0.2010 & 0.1380 & -0.0174 & -0.0127 & 0.0027 \\ 
std & 0.5421 & 0.0792 & 0.3832 & 0.0560 & 0.5231 & 0.0754 & 0.4261 & 0.0622 \\ 
$\wh{\rm std}$ & 0.5527 & 0.0809 & 0.3982 & 0.0582 & 0.5146 & 0.0742 & 0.4213 & 0.0617 \\ 
cvg & 94.8\% & 94.3\% & 5.5\% & 7.7\% & 92.8\% & 93.2\% & 94.4\% & 95.2\% \\ 
\hline
$\tau=0.95$ &&&&&&\\ 
bias & -0.1407 & 0.0126 & 1.6549 & -0.2053 & 0.1307 & -0.0301 & -0.1172 & 0.0038 \\ 
std & 0.6424 & 0.0936 & 0.6407 & 0.0934 & 0.7449 & 0.1043 & 0.4493 & 0.0651 \\ 
$\wh{\rm std}$ & 0.6520 & 0.0958 & 0.6586 & 0.0964 & 0.6947 & 0.0972 & 0.4427 & 0.0645 \\ 
cvg & 95.6\% & 95.7\% & 28.1\% & 42.7\% & 90.8\% & 89.9\% & 94.5\% & 95.0\% \\ 
\end{longtable}

\subsection*{Simulation 2}

The second example differs from the first one only in that the
model error $\epsilon$ is no longer independent of the covariate
$X$. Specifically, we generated $\epsilon$ from a normal distribution
with variance $(0.3 + 0.3X)^2$ and the $\tau$th quantile 0. To adapt
to this scenario, in the
implementation, we slightly modified the constant $c$ in the bandwidth
$h$ to $c=4.0$ for quantiles $0.25, 0.5, 0.75$ and $c=5.0$ for
quantiles $0.05, 0.95$. All other implementation details are
identical to Example 1. We present the corresponding results in
Table \ref{tab:1dep}. Compared to the first example, we observe
that the performance of the proposed and the CLE methods continue
to be satisfactory, while the performance of the Wei-Carroll's method
starts to deteriorate, while still outperforming the naive method.  

\begin{longtable}{cccccccccc}
\caption{Results of Example 2 (Dependent). ``bias'', ``std'', 
  ``$\wh{\rm std}$'', and ``cvg'' are respectively the bias, sample
  standard deviation, estimated standard deviation, the coverage of
  the 95\% confidence intervals of the estimators
  based on 1000 repetitions, and sample size $n=200$.} \label{tab:1dep} \\

\thickhline
Method & \multicolumn{2}{c}{Proposed }&  \multicolumn{2}{c}{Naive }&  \multicolumn{2}{c}{Wei-Carroll's } & \multicolumn{2}{c}{CLE}\\
       & $\beta_0$ & $\beta_1$& $\beta_0$ & $\beta_1$
       & $\beta_0$ & $\beta_1$
       & $\beta_0$ & $\beta_1$\\
\thickhline
\endfirsthead

\multicolumn{10}{c}{{\bfseries \tablename\ \thetable{} -- Continued from previous page}} \\[0.5ex]
\thickhline
Method & \multicolumn{2}{c}{Proposed }&  \multicolumn{2}{c}{Naive }&  \multicolumn{2}{c}{Wei-Carroll's } & \multicolumn{2}{c}{CLE}\\
       & $\beta_0$ & $\beta_1$& $\beta_0$ & $\beta_1$
       & $\beta_0$ & $\beta_1$
       & $\beta_0$ & $\beta_1$\\
\thickhline
\endhead

\hline
\multicolumn{10}{r}{{Continued on next page...}} \\
\endfoot

\thickhline
\endlastfoot

$\tau=0.05$ &&&&&&\\ 
bias & 0.2387 & 0.0635 & 1.3630 & -0.2165 & 0.0029 & 0.0046 & 0.7398 & 0.0850 \\ 
std & 1.7464 & 0.2714 & 1.9857 & 0.3031 & 2.4456 & 0.3804 & 1.7404 & 0.2718 \\ 
$\wh{\rm std}$ & 1.9533 & 0.3043 & 2.1557 & 0.3278 & 2.5280 & 0.3907 & 2.0357 & 0.3257 \\ 
cvg & 95.1\% & 94.2\% & 88.6\% & 90.0\% & 92.5\% & 91.8\% & 94.2\% & 94.8\% \\ 
\hline
$\tau=0.25$ &&&&&&\\ 
bias & 0.1563 & -0.0032 & 1.4462 & -0.2235 & 0.2654 & -0.0398 & 0.1753 & 0.0265 \\ 
std & 1.4501 & 0.2224 & 1.2902 & 0.1955 & 1.5642 & 0.2398 & 1.3739 & 0.2112 \\ 
$\wh{\rm std}$ & 1.5654 & 0.2404 & 1.3882 & 0.2108 & 1.6752 & 0.2559 & 1.4803 & 0.2286 \\ 
cvg & 95.0\% & 95.2\% & 83.2\% & 82.6\% & 94.9\% & 95.0\% & 95.9\% & 95.8\% \\ 
\hline
$\tau=0.5$ &&&&&&\\ 
bias & 0.0770 & -0.0130 & 1.3865 & -0.2112 & 0.1575 & -0.0251 & 0.0901 & -0.0152 \\ 
std & 1.3720 & 0.2098 & 1.2125 & 0.1849 & 1.5215 & 0.2314 & 1.3142 & 0.2014 \\ 
$\wh{\rm std}$ & 1.4584 & 0.2229 & 1.2665 & 0.1929 & 1.5512 & 0.2369 & 1.3869 & 0.2124 \\ 
cvg & 95.0\% & 94.9\% & 79.4\% & 79.4\% & 93.7\% & 94.2\% & 95.5\% & 95.5\% \\ 
\hline
$\tau=0.75$ &&&&&&\\
bias & 0.0447 & -0.0307 & 1.4771 & -0.2226 & 0.1316 & -0.0245 & 0.0164 & -0.0587 \\ 
std & 1.4408 & 0.2206 & 1.2584 & 0.1916 & 1.5853 & 0.2416 & 1.3778 & 0.2120 \\ 
$\wh{\rm std}$ & 1.5730 & 0.2395 & 1.3802 & 0.2094 & 1.7249 & 0.2627 & 1.4753 & 0.2265 \\ 
cvg & 95.2\% & 94.5\% & 82.1\% & 80.8\% & 95.4\% & 95.5\% & 96.2\% & 94.2\% \\ 
\hline
$\tau=0.95$ &&&&&&\\ 
bias & -0.0959 & -0.0895 & 1.5450 & -0.2290 & 0.0457 & -0.0274 & -0.7725 & -0.0784 \\ 
std & 1.7506 & 0.2669 & 2.0018 & 0.3020 & 2.6145 & 0.3886 & 1.8201 & 0.2855 \\ 
$\wh{\rm std}$ & 1.9361 & 0.2972 & 2.1426 & 0.3259 & 2.5386 & 0.3814 & 2.1619 & 0.3482 \\ 
cvg & 95.6\% & 94.0\% & 89.3\% & 89.5\% & 91.9\% & 92.6\% & 95.6\% & 94.8\% \\ 

\end{longtable}

\subsection*{Simulation 3}

Our third simulation study involves a quadratic 
model $Y = \beta_0 + \beta_1 X^2 + \epsilon$, with $(\beta_0,
\beta_1) = (1, 0.3)$, where $\epsilon$ follows
a normal distribution with variance
$(1 - 0.1/|X|)^2$ and $\tau$th quantile 0.
Here the covariate $X$ was generated from the uniform distribution
 on the interval $(0.5, 2.0)$, and all other settings are identical to those in
the second example.

Since both the Wei-Carroll's method and CLE method are
only applicable to linear quantile regression, they are not included in this
example. However, as a ``gold standard'' for comparison we also
calculated the error-free estimator using the true covariate
$X$. Further, we set the bandwidth in the same way as in the previous examples, while the constant $c$ is
 chosen to be $c=11$ for $\tau = 0.5$, $c= 7$ for $\tau =0.25,
 0.75$,  and $c = 3$ for $\tau = 0.05, 0.95$.

 Results from  the proposed, naive and error-free estimators are presented in Table
 \ref{tab:2Quadratic}. As we can see, the performance of the proposed
 method is competitive to that of the error-free estimator,  though it
 has slightly higher standard deviations.
Note that such increased standard deviation is justified and is
  a necessary reflection of the fact that the information is reduced
  when there is measurement error compared to when the covariates are
  precisely observed.
 However, the naive estimator
 performs much worse in the quadratic model than in the linear
 models.

\begin{longtable}{ccccccc}
\caption{Results of  Example 3. ``bias'', ``std'', 
``$\wh{\rm std}$", and ``cvg'' are respectively the bias, sample
standard deviation, estimated standard deviation, the coverage of
the 95\% confidence intervals of the estimators
based on 1000 repetitions, and sample size $n=200$.} \label{tab:2Quadratic} \\

\thickhline
Method & \multicolumn{2}{c}{Error-free } & \multicolumn{2}{c}{Proposed }&  \multicolumn{2}{c}{Naive }\\
       & $\beta_0$ & $\beta_1$ 
       & $\beta_0$ & $\beta_1$ 
       & $\beta_0$ & $\beta_1$\\
\thickhline
\endfirsthead

\multicolumn{7}{c}{{\bfseries \tablename\ \thetable{} -- Continued from previous page}} \\[0.5ex]
\thickhline
Method & \multicolumn{2}{c}{Error-free } & \multicolumn{2}{c}{Proposed }&  \multicolumn{2}{c}{Naive }\\
       & $\beta_0$ & $\beta_1$ 
       & $\beta_0$ & $\beta_1$ 
       & $\beta_0$ & $\beta_1$\\
\thickhline
\endhead

\hline
\multicolumn{7}{r}{{Continued on next page...}} \\
\endfoot

\thickhline
\endlastfoot

$\tau=0.05$ &&&&&&\\ 
bias & 0.0116 & -0.0059 & -0.0322 & 0.0373 & 0.2455 & -0.1875  \\ 
std & 0.2329 & 0.1191 & 0.2733 & 0.1390 & 0.1961 & 0.0745  \\ 
$\wh{\rm std}$ & 0.2756 & 0.1406 & 0.2506 & 0.1480 & 0.2166 & 0.0860  \\ 
cvg & 95.3\% & 97.7\% & 94.1\% & 96.1\% & 76.4\% & 36.5\%  \\ 
\hline
$\tau=0.25$ &&&&&&\\ 
bias & 0.0136 & -0.006 & -0.0885 & 0.0126 & 0.2712 & -0.1885  \\ 
std & 0.1559 & 0.0783 & 0.2118 & 0.1175 & 0.1298 & 0.0503  \\ 
$\wh{\rm std}$ & 0.1701 & 0.0865 & 0.2165 & 0.1236 & 0.1396 & 0.0560  \\ 
cvg & 95.2\% & 96.0\% & 96.4\% & 96.4\% & 48.8\% & 7.6\%  \\ 
\hline
$\tau=0.5$ &&&&&&\\ 
bias & 0.0052 & -0.0043 & -0.0106 & 0.0039 & 0.2834 & -0.1855  \\ 
std & 0.1458 & 0.0748 & 0.2038 & 0.1145 & 0.1207 & 0.0469  \\ 
$\wh{\rm std}$ & 0.1553 & 0.0792 & 0.2162 & 0.1218 & 0.1297 & 0.0524  \\ 
cvg & 95.4\% & 95.6\% & 96.0\% & 95.0\% & 41.5\% & 5.2\%  \\ 
\hline
$\tau=0.75$ &&&&&&\\
bias & 0.0016 & -0.0066 & 0.0721 & -0.0105 & 0.2983 & -0.1823  \\ 
std & 0.1564 & 0.0782 & 0.2017 & 0.1080 & 0.1315 & 0.0531  \\ 
$\wh{\rm std}$ & 0.1679 & 0.0853 & 0.2111 & 0.1147 & 0.1453 & 0.0589  \\ 
cvg & 95.1\% & 95.7\% & 94.0\% & 95.3\% & 45.2\% & 14.3\%  \\ 
\hline
$\tau=0.95$ &&&&&&\\ 
bias & 0.0064 & -0.0085 & -0.0043 & -0.0271 & 0.3301 & -0.1816  \\ 
std & 0.2587 & 0.1263 & 0.2561 & 0.1182 & 0.2303 & 0.0909  \\ 
$\wh{\rm std}$ & 0.2782 & 0.1415 & 0.2408 & 0.1265 & 0.2465 & 0.1012  \\ 
cvg & 94.5\% & 95.5\% & 93.9\% & 96.0\% & 74.0\% & 52.8\%  \\ 

\end{longtable}

\subsection*{Simulation 4}

We next conducted a fourth simulation study where
the quantile regression model has the form
$Y = \beta_0 + \beta_1 \sin(\beta_2 X) + \epsilon$. 
We set $(\beta_0, \beta_1, \beta_2) = (-1, 1, 1)$,
 generated $X$ from Uniform$(0.2\pi, 1.2\pi)$ 
and $\epsilon$ from a normal distribution with variance
$(0.8 + 0.3|\sin(X)|)^2$ and quantile 0. The measurement error has
standard deviation 0.25. In implementation, we set the constant $c$ in
the bandwidth expression at $c=4.5$ for all $\tau =0.05, 0.25, 0.5,
0.75, 0.95$. We used a sample size $n=500$ in this example
and performed 200 bootstrap samples to assess the variability of the estimators. 

The results are presented in Table \ref{tab:3dep}.
Again, our proposed method performed very competitively with the
error-free estimator, except for the intercept parameter in the case
of $\tau=0.05$ and $n=500$. We suspect this is a finite sample
performance issue and hence increased the sample size to $n=2000$ and
$n= 5000$ for $\tau=0.05$.  
Indeed, the standard errors continuously decreased and the coverage
probabilities became closer to the nominal level of 95\%. In all the
cases, the naive estimator shows persistently very large bias and
subsequently invalid confidence intervals.  

\begin{longtable}{cccccccccc}
\caption{Results of  Example  4. ``bias'', ``std'', 
  ``$\wh{\rm std}$", and ``cvg'' are respectively the bias, sample
  standard deviation, estimated standard deviation, the coverage of
  the 95\% confidence intervals of the estimators
  based on 1000 repetitions.} \label{tab:3dep} \\

\thickhline
Method & \multicolumn{3}{c}{Error-free } & \multicolumn{3}{c}{Proposed }&  \multicolumn{3}{c}{Naive }\\
       & $\beta_0$ & $\beta_1$ & $\beta_2$ 
       & $\beta_0$ & $\beta_1$ & $\beta_2$
       & $\beta_0$ & $\beta_1$ & $\beta_2$   \\
\thickhline
\endfirsthead

\multicolumn{10}{c}{{\bfseries \tablename\ \thetable{} -- Continued from previous page}} \\[0.5ex]
\thickhline
Method & \multicolumn{3}{c}{Error-free } & \multicolumn{3}{c}{Proposed }&  \multicolumn{3}{c}{Naive }\\
       & $\beta_0$ & $\beta_1$ & $\beta_2$ 
       & $\beta_0$ & $\beta_1$ & $\beta_2$
       & $\beta_0$ & $\beta_1$ & $\beta_2$   \\
\thickhline
\endhead

\hline
\multicolumn{10}{r}{{Continued on next page...}} \\
\endfoot

\thickhline
\endlastfoot

$\tau=0.05$ &$n=500$&&&&&\\ 
bias & -0.0068 & 0.0228 & 0.0145 & 0.1139 & 0.0581 & 0.0101 & 0.1757 & -0.1917 & 0.0952  \\ 
std & 0.2400 & 0.2475 & 0.0852 & 0.2898 & 0.3099 & 0.0973 & 0.1755 & 0.1679 & 0.0935  \\ 
$\wh{\rm std}$ & 0.2491 & 0.2598 & 0.0892 & 0.3191 & 0.3589 & 0.1069 & 0.1859 & 0.1852 & 0.0956  \\ 
cvg & 91.9\% & 94.3\% & 94.6\% & 89.9\% & 97.4\% & 96.6\% & 74.9\% & 72.9\% & 82.5\%  \\ 
\hline
$\tau=0.05$ &$n$=2000&&&&&\\ 
bias & -0.0062 & 0.0113 & 0.0075 & 0.0745 & 0.0374 & 0.0032 & 0.1793 & -0.2026 & 0.0891  \\ 
std & 0.1704 & 0.1774 & 0.0617 & 0.2182 & 0.2386 & 0.0746 & 0.1199 & 0.1165 & 0.0671  \\ 
$\wh{\rm std}$ & 0.1780 & 0.1865 & 0.0633 & 0.2402 & 0.2714 & 0.0812 & 0.1296 & 0.1285 & 0.069   \\ 
cvg & 93.1\% & 93.7\% & 94.9\% & 92.5\% & 97.5\% & 96.0\% & 63.6\% & 57.7\% & 76.5\%  \\ 
\hline    
$\tau=0.05$ &$n$=5000&&&&&\\ 
bias & -0.0082 & 0.0118 & 0.0019 & 0.0236 & 0.0451 & -0.0037 & 0.1821 & -0.2079 & 0.0863  \\ 
std & 0.1146 & 0.1209 & 0.0390 & 0.1682 & 0.1916 & 0.0562 & 0.0784 & 0.0777 & 0.0432  \\ 
$\wh{\rm std}$ & 0.1148 & 0.1193 & 0.0400 & 0.1645 & 0.1924 & 0.0598 & 0.0811 & 0.0797 & 0.0443  \\ 
cvg & 93.1\% & 93.5\% & 93.3\% & 95.3\% & 93.6\% & 95.2\% & 40.1\% & 28.5\% & 52.0\%  \\ 
\thickhline  
$\tau=0.25$ &$n=500$&&&&&\\ 
bias & -0.0045 & 0.0106 & 0.0076 & 0.0362 & 0.0166 & 0.0091 & 0.1841 & -0.2003 & 0.0812  \\ 
std & 0.1586 & 0.1612 & 0.0565 & 0.1788 & 0.1843 & 0.0640 & 0.1128 & 0.104 & 0.0608  \\ 
$\wh{\rm std}$ & 0.1625 & 0.1685 & 0.0575 & 0.1998 & 0.2159 & 0.0667 & 0.1177 & 0.1164 & 0.0624  \\ 
cvg & 93.4\% & 93.8\% & 93.9\% & 93.1\% & 95.0\% & 95.5\% & 60.4\% & 55.2\% & 76.2\%  \\ 
\thickhline      
$\tau=0.5$ &$n=500$&&&&&\\ 
bias & -0.0078 & 0.0125 & 0.0051 & -0.0144 & 0.0227 & 0.0059 & 0.1869 & -0.2023 & 0.0736  \\ 
std & 0.1489 & 0.1543 & 0.0518 & 0.1704 & 0.1823 & 0.0583 & 0.1050 & 0.1048 & 0.0553  \\ 
$\wh{\rm std}$ & 0.1517 & 0.1582 & 0.0535 & 0.1890 & 0.2046 & 0.0623 & 0.1101 & 0.1101 & 0.0581  \\ 
cvg & 94.7\% & 94.6\% & 95.0\% & 96.0\% & 95.2\% & 95.3\% & 55.5\% & 50.4\% & 79.0\%  \\ 
\thickhline
$\tau=0.75$ &$n=500$&&&&&\\
bias & -0.0051 & 0.0101 & 0.0075 & -0.0632 & 0.0283 & 0.0078 & 0.1914 & -0.2033 & 0.0711  \\ 
std & 0.1598 & 0.1648 & 0.0566 & 0.1930 & 0.2088 & 0.0664 & 0.1124 & 0.1133 & 0.0607  \\ 
$\wh{\rm std}$ & 0.1642 & 0.1710 & 0.0581 & 0.2149 & 0.2354 & 0.0703 & 0.1198 & 0.1205 & 0.0631  \\ 
cvg & 94.1\% & 94.1\% & 95.1\% & 95.7\% & 94.5\% & 95.2\% & 59.2\% & 56.7\% & 82.1\%  \\ 
\thickhline
$\tau=0.95$ &$n=500$&&&&&\\ 
bias & -0.0210 & 0.0325 & 0.0146 & -0.1404 & -0.0109 & -0.0129 & 0.1768 & -0.1798 & 0.0642  \\ 
std & 0.2465 & 0.2561 & 0.0881 & 0.3598 & 0.4478 & 0.1414 & 0.1749 & 0.1781 & 0.0902  \\ 
$\wh{\rm std}$ & 0.2464 & 0.2571 & 0.0902 & 0.3559 & 0.4404 & 0.1492 & 0.1856 & 0.1905 & 0.0963  \\ 
cvg & 93.1\% & 93.9\% & 92.1\% & 95.9\% & 94.8\% & 94.9\% & 79.3\% & 78.4\% & 90.4\%  \\ 
\thickhline
\end{longtable}

\subsection*{Simulation 5}

Following the request of a referee, we conducted an additional
simulation study which includes precisely measured covariates, and
a systematic bandwidth selection procedure.
The  data generation mimics the read data analyzed
in Section \ref{sec:real}.
Specifically, we consider the quantile regression model 
\bse
Y = 100\beta_0 +  \beta_1 X_{1} +10\beta_2 \log(X_{2}) +
\beta_3 Z_{1}  + \beta_4 Z_{2} + \beta_5 Z_{3} +  \epsilon, 
\ese 
with $(\beta_0, \beta_1, \beta_2, \beta_3, \beta_4, \beta_5) =
(1.5, -0.7, -3.2, -0.1, 0.7, -0.1)$. Here,
the error $\epsilon$ follows a normal distribution with its $\tau$th
quantile zero and a heteroscedastic variance $(2.0 +
X_{2}^{-0.1})^2$. 
The five covariates were generated from the following distributions:
\begin{itemize}
\item $Z_1=R_bR_g$, where $R_b$ is
a Bernoulli random 
variable with success probability 0.6 and $R_g$ is a Gamma random variable
following
$\text{Gamma}(\alpha = 0.56, \beta = 0.06)$;
\item $(X_1, Z_{2}, Z_{3})\trans = \{6+3R_1, 36+2R_2, 130 + 14\Phi(R_3)\}\trans$
where $\Phi(\cdot)$ is the standard normal cdf and
\bse
\begin{pmatrix}R_1\\
R_2\\
R_3
\end{pmatrix} \sim \MVN
\begin{Bmatrix}
\begin{pmatrix}
0\\
0\\
0
\end{pmatrix}, \,
\left(
\begin{array}{rrr} 
1.0 & -0.2 & 0.0 \\
-0.2 & 1.0 & 0.7 \\
0.0 & 0.7 & 1.0
\end{array}
\right)
\end{Bmatrix};\\
\ese
\item $X_2$ follows $N(60, 13^2)$.
\end{itemize}
Further, we generated the two
measurement errors independently from $U_{1}\sim N(0, 3.88)$,
$U_{2}\sim N(0, 1.85)$, and formed $W_k=X_k+U_k$ for $k=1, 2$.
A total of $n=1000$ independent observations 
were generated.

In implementing the method for parameter estimation,  we set
  $c=12.0$ for $\tau=0.5$, 
  $c=13.0$ for $\tau=0.25, 0.75$, and $c=18.5$ for $\tau=0.05, 0.95$
  following a 10-fold CV procedure with
  $\lambda$ from 0 to 2 by a step size 0.25 and $B=20$, see
  details of the bandwidth selection procedure in Section
  \ref{sec:bandwidth}. We further estimated 
  the variability of the estimators via 200 bootstrap samples. 
  
Tables \ref{tab:mimic_X} and \ref{tab:mimic_Z} summarize the
performance of the proposed, error-free, and naive estimators.
We find that the proposed estimators perform remarkably similar to the
error-free estimators. Both demonstrate minimal bias and maintain
coverage probabilities near the nominal level.
In contrast, the naive estimators, which ignore the measurement error
in $\W$, exhibit substantial bias and a significant degradation in
coverage probability, particularly for the coefficients $(\beta_0,
\beta_1, \beta_2)\trans$. 
It is interesting to note that the naive estimates of $\beta_4,
\beta_5$ also have substantial bias and deteriorated coverage
probability due the correlation of $Z_{2}, Z_{3}$ with $X_1$ even
though they are measured precisely.
Lastly, due to the limited sample size and large portion of zeros
values in $Z_1$, the inference results its coefficient  $\beta_3$  is generally
worse than other parameters, even for the
error-free case.

 \begin{longtable}{cccccccccc}
\caption{Results for $(\beta_0,\, \beta_1,\, \beta_2)^{\top}$ of  Example  5. ``bias'', ``std'', 
   ``$\wh{\rm std}$", and ``cvg'' are respectively the bias, sample
   standard deviation, estimated standard deviation, the coverage of
   the 95\% confidence intervals of the estimators
   based on 1000  repetitions, and sample size $n=1000$.} \label{tab:mimic_X} \\

\thickhline
Method & \multicolumn{3}{c}{Error-free } & \multicolumn{3}{c}{Proposed }&  \multicolumn{3}{c}{Naive }\\
       & $\beta_0$ & $\beta_1$ & $\beta_2$ 
       & $\beta_0$ & $\beta_1$ & $\beta_2$
       & $\beta_0$ & $\beta_1$ & $\beta_2$   \\
\thickhline
\endfirsthead

\multicolumn{10}{c}{{\bfseries \tablename\ \thetable{} -- Continued from previous page}} \\[0.5ex]
\thickhline
Method & \multicolumn{3}{c}{Error-free } & \multicolumn{3}{c}{Proposed }&  \multicolumn{3}{c}{Naive }\\
       & $\beta_0$ & $\beta_1$ & $\beta_2$ 
       & $\beta_0$ & $\beta_1$ & $\beta_2$
       & $\beta_0$ & $\beta_1$ & $\beta_2$   \\
\thickhline
\endhead

\hline
\multicolumn{10}{r}{{Continued on next page...}} \\
\endfoot

\thickhline
\endlastfoot

$\tau=0.05$ &&&&&&\\ 
bias & -0.0029 & 0.0014 & 0.0066 & 0.0003 & -0.0020 & -0.0055 & -0.0416 & 0.2195 & 0.0876  \\ 
std & 0.0898 & 0.1124 & 0.2056 & 0.0215 & 0.0491 & 0.0507 & 0.1079 & 0.0876 & 0.2488  \\ 
$\wh{\rm std}$ & 0.0344 & 0.0707 & 0.0846 & 0.0216 & 0.0501 & 0.0528 & 0.0893 & 0.0824 & 0.2103  \\ 
cvg & 96.1\% & 94.8\% & 96.1\% & 94.8\% & 94.7\% & 95.4\% & 94.9\% & 18.3\% & 96.3\%\\ 
\hline
$\tau=0.25$ &&&&&&\\ 
bias & -0.0007 & 0.0028 & 0.0017 & -0.0008 & 0.0094 & 0.0008 & -0.0254 & 0.2221 & 0.0568  \\ 
std & 0.0206 & 0.0388 & 0.0503 & 0.0203 & 0.0495 & 0.0499 & 0.0227 & 0.0365 & 0.0555  \\ 
$\wh{\rm std}$ & 0.0212 & 0.0424 & 0.0521 & 0.0207 & 0.0464 & 0.0506 & 0.0254 & 0.0407 & 0.0621  \\ 
cvg & 95.2\% & 95.4\% & 95.2\% & 95.9\% & 94.8\% & 96.0\% & 85.4\% & 0.1\% & 87.7\%\\ 
\hline   
$\tau=0.5$ &&&&&&\\ 
bias & -0.0003 & 0.0017 & 0.0006 & -0.0006 & 0.0126 & 0.0015 & -0.0159 & 0.2233 & 0.0390  \\ 
std & 0.0191 & 0.0359 & 0.0469 & 0.0197 & 0.0443 & 0.0482 & 0.0208 & 0.0335 & 0.0507  \\ 
$\wh{\rm std}$ & 0.0193 & 0.0387 & 0.0474 & 0.0199 & 0.0441 & 0.0487 & 0.0234 & 0.0370 & 0.0572  \\ 
cvg & 94.3\% & 95.3\% & 94.3\% & 95.1\% & 94.7\% & 95.1\% & 91.5\% & 0.1\% & 91.5\%\\ 
\hline
$\tau=0.75$ &&&&&&\\
bias & -0.0004 & 0.0020 & 0.0009 & -0.0003 & 0.0094 & 0.0021 & -0.0065 & 0.2233 & 0.0212  \\ 
std & 0.0201 & 0.0388 & 0.0490 & 0.0200 & 0.0482 & 0.0491 & 0.0223 & 0.0367 & 0.0544  \\ 
$\wh{\rm std}$ & 0.0209 & 0.0423 & 0.0514 & 0.0203 & 0.0464 & 0.0496 & 0.0254 & 0.0410 & 0.0622  \\ 
cvg & 94.2\% & 95.3\% & 94.4\% & 94.4\% & 95.0\% & 94.2\% & 96.3\% & 0.1\% & 95.8\%\\ 
\hline
$\tau=0.95$ &&&&&&\\ 
bias & -0.0019 & 0.0014 & 0.0051 & -0.0008 & -0.0009 & 0.0065 & 0.0049 & 0.2221 & 0.0019  \\ 
std & 0.0504 & 0.0729 & 0.1316 & 0.0220 & 0.0480 & 0.0517 & 0.0577 & 0.1236 & 0.1570  \\ 
$\wh{\rm std}$ & 0.0346 & 0.0755 & 0.0849 & 0.0210 & 0.0502 & 0.0514 & 0.0524 & 0.0845 & 0.1318  \\ 
cvg & 94.3\% & 96.1\% & 94.2\% & 94.1\% & 96.1\% & 94.2\% & 97.9\% & 16.1\% & 98.3\%\\ 
\thickhline
\end{longtable}
 
 \begin{longtable}{cccccccccc}
\caption{Results  for $(\beta_3,\, \beta_4,\, \beta_5)^{\top}$ of  Example  5. ``bias'', ``std'', 
    ``$\wh{\rm std}$", and ``cvg'' are respectively the bias, sample
    standard deviation, estimated standard deviation, the coverage of
    the 95\% confidence intervals of the estimators
    based on 1000  repetitions, and sample size $n=1000$.} \label{tab:mimic_Z} \\

\thickhline
Method & \multicolumn{3}{c}{Error-free } & \multicolumn{3}{c}{Proposed }&  \multicolumn{3}{c}{Naive }\\
       & $\beta_3$ & $\beta_4$ & $\beta_5$ 
       & $\beta_3$ & $\beta_4$ & $\beta_5$
       & $\beta_3$ & $\beta_4$ & $\beta_5$   \\
\thickhline
\endfirsthead

\multicolumn{10}{c}{{\bfseries \tablename\ \thetable{} -- Continued from previous page}} \\[0.5ex]
\thickhline
Method & \multicolumn{3}{c}{Error-free } & \multicolumn{3}{c}{Proposed }&  \multicolumn{3}{c}{Naive }\\
       & $\beta_3$ & $\beta_4$ & $\beta_5$ 
       & $\beta_3$ & $\beta_4$ & $\beta_5$
       & $\beta_3$ & $\beta_4$ & $\beta_5$   \\
\thickhline
\endhead

\hline
\multicolumn{10}{r}{{Continued on next page...}} \\
\endfoot

\thickhline
\endlastfoot

$\tau=0.05$ &&&&&&\\ 
bias & 0.0001 & -0.0221 & 0.0065 & -0.0003 & 0.0073 & -0.0042 & 0.0007 & 0.0876 & -0.0324  \\ 
std & 0.0221 & 0.1862 & 0.0771 & 0.0139 & 0.1831 & 0.0998 & 0.0256 & 0.2067 & 0.1028  \\ 
$\wh{\rm std}$ & 0.0163 & 0.1849 & 0.0819 & 0.0113 & 0.0945 & 0.0466 & 0.0243 & 0.2637 & 0.1182  \\ 
cvg & 88.8\% & 94.1\% & 95.9\% & 93.8\% & 94.8\% & 95.0\% & 95.3\% & 97.2\% & 97.5\%\\ 
\hline
$\tau=0.25$ &&&&&&\\ 
bias & -0.0001 & 0.0073 & -0.0027 & -0.0002 & 0.0090 & -0.0030 & 0.0000 & 0.1266 & -0.0427  \\ 
std & 0.0110 & 0.0875 & 0.0422 & 0.0107 & 0.0762 & 0.0415 & 0.0124 & 0.0955 & 0.0468  \\ 
$\wh{\rm std}$ & 0.0107 & 0.0985 & 0.0460 & 0.0103 & 0.0790 & 0.0383 & 0.0133 & 0.1268 & 0.0568  \\ 
cvg & 90.8\% & 95.8\% & 95.3\% & 93.9\% & 95.6\% & 94.5\% & 94.7\% & 87.3\% & 92.7\%\\ 
\hline  
$\tau=0.5$ &&&&&&\\ 
bias & -0.0003 & 0.0051 & -0.0019 & -0.0005 & 0.0111 & -0.0037 & -0.0006 & 0.1266 & -0.0428  \\ 
std & 0.0103 & 0.0796 & 0.0389 & 0.0101 & 0.0740 & 0.0390 & 0.0112 & 0.0857 & 0.0426  \\ 
$\wh{\rm std}$ & 0.0100 & 0.0861 & 0.0405 & 0.0100 & 0.0765 & 0.0369 & 0.0120 & 0.1082 & 0.0493  \\ 
cvg & 90.7\% & 96.0\% & 95.3\% & 93.7\% & 95.5\% & 94.9\% & 96.1\% & 80.6\% & 88.8\%\\ 
\hline
$\tau=0.75$ &&&&&&\\
bias & -0.0004 & 0.0042 & -0.0008 & -0.0006 & 0.0092 & -0.0024 & -0.0011 & 0.1282 & -0.0416  \\ 
std & 0.0107 & 0.0835 & 0.0418 & 0.0118 & 0.0769 & 0.0407 & 0.0125 & 0.0936 & 0.0475  \\ 
$\wh{\rm std}$ & 0.0109 & 0.0953 & 0.0447 & 0.0104 & 0.0790 & 0.0381 & 0.0132 & 0.1216 & 0.0550  \\ 
cvg & 92.4\% & 97.0\% & 95.4\% & 93.0\% & 94.7\% & 93.8\% & 95.4\% & 85.6\% & 90.0\%\\ 
\hline
$\tau=0.95$ &&&&&&\\ 
bias & 0.0001 & -0.0219 & 0.0063 & 0.0000 & 0.0023 & -0.0014 & 0.0006 & 0.0883 & -0.0328  \\ 
std & 0.0220 & 0.1857 & 0.0769 & 0.0109 & 0.0812 & 0.0406 & 0.0256 & 0.2064 & 0.1028  \\ 
$\wh{\rm std}$ & 0.0163 & 0.1849 & 0.0820 & 0.0106 & 0.0830 & 0.0399 & 0.0243 & 0.2635 & 0.1183  \\ 
cvg & 89.0\% & 94.2\% & 96.0\% & 93.9\% & 94.8\% & 95.0\% & 95.4\% & 97.2\% & 97.5\%\\ 
\thickhline
\end{longtable}

\section{Real data application} \label{sec:real}
\subsection{Data structure}
We used the proposed method to analyze a subset of the data from
Kaggle, Japan Cherry Blossoms Forecasts
2024\footnote{https://www.kaggle.com/datasets/altabbt/japan-cherry-blossoms-forecasts-2024}. The
complete  weather data set was downloaded from
Open-meteo\footnote{https://open-meteo.com/en/docs/historical-weather-api}.  
The analyzed data spanned from February 29 to March 18, 2024, yielding 13,794 observations across 903 locations in Japan.
At each location $i$,  the response variable $Y_i$ is defined as the number of days from the recorded date to
\texttt{Kaika} (blossoming) which represents the wait time until blossom. 
 We included five other variables as covariates, among which
the daily average temperature ($X_{1i}$) and the growth rate of cherry blossom buds (\texttt{meter}, $X_{2i}$) are considered two important factors  affecting the enzyme activity and reflecting plant growth status. 
However, these two variables contain measurement errors and so we can only observe their surrogates $W_{1i}$ and $W_{2i}$ respectively.
The remaining three covariates are daily precipitation in total
including snow ($Z_{1i}$), the latitude of the location ($Z_{2i}$),
and the longitude of the location ($Z_{3i}$).
We randomly selected a one-day record for each location and considered
the following nonlinear quantile regression model at $\tau=0.5$ with
measurement error in recorded daily average temperature and
\texttt{meter}, 
\be \label{model:realdata}
Y_i &=& 100\cdot\beta_0 +  \beta_1 X_{1i} +10\cdot\beta_2 \log(X_{2i}) + \beta_3 Z_{1i}  + \beta_4 Z_{2i} + \beta_5 Z_{3i} +  \epsilon,\\
W_{1i}&=& X_{1i} + U_{1i},\n\\
W_{2i}&=& X_{2i} + U_{2i},\n
\ee 
where $\bb = (\beta_0, \beta_1, \beta_2, \beta_3, \beta_4, \beta_5)\trans$, $U_{1i} \sim N(0, \sigma_1^2)$, $U_{2i} \sim N(0, \sigma_2^2)$, and $i=1, ..., 903$.

\subsection{Implementation}
We implemented the proposed method as well as the naive method that
estimates parameters using the observed data directly. In order to
obtain an estimate for the measurement error covariance matrix $\bSigma$, 
we adopt an interpolation method using the data 
$(\W_{ij}, \t_j)$, $j=1, ..., m_i$, $i = 1, .., n$, where $t_j$ is the
time when $W_{ij}$ is recorded and $m_i$ is the number of days on
which the data was collected. Specifically, for each $i=1,2,...,n$, we
fit a simple linear model for the daily average temperature and a
Poisson model for the \texttt{meter} as
\bse
X_{1ij} &=&  a_{1i} +  b_{1i} t_j  + \epsilon_{1i},\\
X_{2ij}  &=&  \exp(a_{2i} +  b_{2i} t_j) + \epsilon_{2i},
\ese
where $a_{1i}$, $a_{2i}$, and $b_{1i}$, $b_{2i}$ are parameters. Then
the estimated equations are used to estimate the true covariates as
$\wh x_{1i} =\wh a_{1i} +  \wh b_{1i} t_{k_i}$ and $\wh x_{2i}
=\exp(\wh a_{2i} +  \wh b_{2i} t_{k_i})$, for a randomly selected
$k_i\in\{1,..., m_i\}$. This analysis produces consistent trends for
individual covariates, thus, reducing the measurement errors and can
be viewed as a version of the estimator when there is no
measurement error. 
Similar practice has been employed in \cite{terry2007maternal} and \cite{wei2009quantile}.
Using the interpolated values $\wh\x_i$, we obtain the estimate of $\bSigma$ as
\bse 
\wh\bSigma = \frac{1}{903}\sum_{i=1}^{903}\frac{1}{m_i-2}
\begin{pmatrix}
(\w_{1i} - \wh\x_{1i})\trans(\w_{1i} - \wh\x_{1i}) & 0 \\
0 & (\w_{2i} - \wh\x_{2i})\trans(\w_{2i} - \wh\x_{2i})
\end{pmatrix} = 
\begin{pmatrix}
3.88 & 0 \\
0 & 1.85
\end{pmatrix}.
\ese
Further, similar to the previous examples, we approximated the
integrals in the estimating equation \eqref{eq:esteqc} using a
two-dimensional Gauss-Hermite quadrature and the bandwidth tuning
parameter $c=15$, 
and obtained the variance of the estimators by 200 bootstrap samples. Finally, as a by-product and an alternative method,
we also used the estimated data $\wh\x_i$, $i=1,2,...,n$, to estimate
the above quantile regression model similar to the error-free
estimator in Example 3 and 4.

\subsection{Results}
\begin{figure}[H]
\centering
\includegraphics[width=0.95\textwidth]{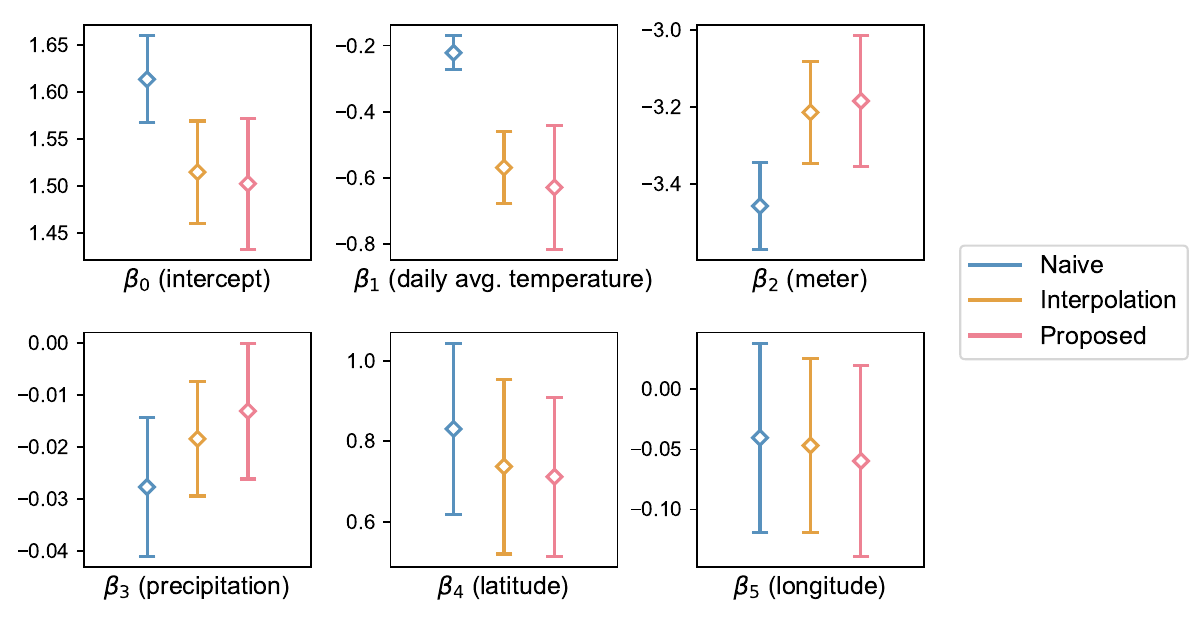}
\caption{Comparison of different methods. $\Diamond$ and lines
    represent the estimator and its 95\% confidence
    interval.}\label{fig:realdata}
\end{figure}

The estimation results are given in Table \ref{tab:realdata} and Figure \ref{fig:realdata}. 
Overall, both the interpolation and our proposed method produced
similar estimates, while the proposed estimator has larger estimated
standard deviation, reflecting the information loss due to the
  presence of measurement errors. All estimated values for coefficients
$\beta_1,\beta_2,\beta_3,\beta_4$ have the correct signs indicating
that the median wait time until blossom deceases with the increase of
the daily average temperature ($X_{1i}$), the \texttt{meter}
($X_{2i}$), and the daily precipitation ($Z_{1i}$) respectively, while
increases with the increase of the latitude of the location ($Z_{2i}$)
due to weather conditions. Further, both methods produced high
p-values for $\beta_5$ indicating that the longitude of the location
($Z_{3i}$) is statistically insignificant. Finally, the naive
estimates are notably different from the two methods, particularly it
significantly underestimates the effect of the daily average
temperature which is the main determinant for the response variable.

\begin{longtable}{lllllll}
\caption{Results of the real data application. ``est'', ``$\wh{\rm
    std}$'' and ``$p$-value'' are respectively the estimate, estimated
  standard deviation and $p$-value.}\label{tab:realdata} \\

\thickhline
&$\beta_0$ & $\beta_1$ & $\beta_2$ & $\beta_3$ & $\beta_4$ & $\beta_5$ \\
\thickhline
\endfirsthead

\multicolumn{7}{c}{{\bfseries \tablename\ \thetable{} -- Continued from previous page}} \\[0.5ex]
\thickhline
&$\beta_0$ & $\beta_1$ & $\beta_2$ & $\beta_3$ & $\beta_4$ & $\beta_5$ \\
\thickhline
\endhead

\hline
\multicolumn{7}{r}{{Continued on next page...}} \\
\endfoot

\thickhline
\endlastfoot

est&  &  &  &  &  &   \\
Naive & 1.6138 & -0.2219 & -3.4568 & -0.0277 & 0.8306 & -0.0403 \\
Interpolation & 1.5151 & -0.5688 & -3.2135 & -0.0184 & 0.7371 & -0.0468 \\
Proposed & 1.5028 & -0.6285 & -3.1840 & -0.0131 & 0.7117 & -0.0597 \\\hline
\\ [-1em]
$\wh{\rm std}$&  &  &  &  &  &   \\
Naive & 0.0236 & 0.0262 & 0.0572 & 0.0068 & 0.1088 & 0.0402 \\
Interpolation & 0.0277 & 0.0559 & 0.0676 & 0.0056 & 0.1108 & 0.0369 \\
Proposed & 0.0355 & 0.0951 & 0.0864 & 0.0066 & 0.1009 & 0.0405 \\\hline
$p$-value&  &  &  &  &  &   \\
Naive         & 0.0 & 0.0 & 0.0 & 4.5e-05 & 2.2e-14 & 0.3159 \\
Interpolation & 0.0 & 0.0 & 0.0 & 9.7e-04 & 2.9e-11 & 0.2044 \\
Proposed      & 0.0 & 3.9e-11 & 0.0 & 0.0486 & 1.7e-12 & 0.1405 \\

\end{longtable}

\section{Discussion}
Although the quantile regression method has a lot of advantages and
desired properties over the mean regression method in real data
analysis, it entails substantial technical difficulties due to the
mathematical complexity of the quantile function. In addition,
incorporating the measurement error in covariates in general nonlinear
quantile regression poses extra challenges and requires more
sophisticated mathematical tools.
We have developed a consistent estimation approach for general quantile
regression with measurement error. To our best knowledge, this is the
first consistent estimator in nonlinear quantile regression with
measurement error without imposing assumptions on other quantile
levels that are not under study. We foresee that the filling of this
literature gap opens
the door of quantile regression in many real data applications where
covariates are subject to measurement errors. 
Interestingly, our method can also be used for quantile regression when
the response variable has normal additive error. These problems
in a more general context are studied
in \cite{hausman2021errors}. 
We point out that our method is applicable only when the measurement
error is normal additive.
Development in measurement error problems 
has encompassed non-normal or even non-additive structures for 
parametric regression \citep{tsiatisma2004, matsiatis2006} 
or mean regression models \citep{garciama2017} 
without relying on deconvolution procedures. 
How to incorporate 
more flexible measurement error structures 
while still bypassing
 the need to estimate the distribution of error-prone covariates is a
challenging yet interesting future research topic.

\newpage
\section*{Appendix}
\addcontentsline{toc}{section}{Appendices}
\renewcommand{\thesubsection}{\Alph{subsection}}

\subsection{Proof of Theorem \ref{th:classic}} \label{apdx:th1pf}
We first show that $\wh\bb_s$ is a consistent estimator of $\bb$
by using Theorem 2.1 of \cite{newey1994large}.
We regard finding the solution of $n^{-1}\sumi \S(Y_i,\W_i,\Z_i, \bb, \bSigma)\}=\0$
as a maximization problem with objective function
$\wh Q_n(\bb)\equiv -\| n^{-1}\sumi\S(Y_i,\W_i,\Z_i, \bb, \bSigma)\}
\|_2^2$. 
Define $Q_0(\bb) \equiv -\| E\{\S(Y,\W,\Z, \bb, \bSigma)\} \|_2^2$. Under
  Condition \ref{con:uniroot}, $Q_0(\bb)$ is uniquely maximized at the true
  $\bb$ in its neighborhood. Conditions \ref{con:compact} and
  \ref{con:bddregfunc} ensure that $\calB$ is compact and $Q_0(\bb)$
  is continuous on $\calB$. Lastly, $\wh Q_n(\bb)\to Q_0(\bb)$
  uniformly  in probability by the law of large numbers and the compactness of $\calB$. 
 Therefore, the consistency of $\wh\bb_s$ follows.

   Next, we find the asymptotic distribution for the
estimator $\wh\bb_s$. 
A Taylor expansion leads to
\bse
\0&=& n^{-1/2}\sumi   \S(Y_i, \W_i, \Z_i, \wh\bb_s, \bSigma)\\
&=&  n^{-1/2}\sumi \S(Y_i, \W_i, \Z_i, \bb, \bSigma)
+\frac{1}{n}\sumi \frac{\partial\,   \S(Y_i, \W_i, \Z_i, \bb, \bSigma)}{\partial\bb\trans}\bigg\rvert_{\bb = \bb^*}\sqrt{n}(\wh\bb_s-\bb),
\ese
 where $\bb^*$ is on the line connecting $\wh\bb_s$ and $\bb$.
Now
\be\label{eq:ac}
&& E\left\{\frac{ \partial   \S(Y, \W, \Z, \bb, \bSigma)}{\partial\bb\trans}\right\}\n\\
&=&  -2 E\left[
k_h\left\{Y-m(\W+\sqrt{-1}\bSigma^{1/2}\V,\Z,\bb)\right\}
\m_\bb'(\W+\sqrt{-1}\bSigma^{1/2}\V,\Z,\bb)^{\otimes2}
\right]\n\\
&&\textstyle 
 -E\left[\frac{Y-m(\W+\sqrt{-1}\bSigma^{1/2}\V,\Z,\bb)}{h^2}k'\left\{\frac{Y-m(\W+\sqrt{-1}\bSigma^{1/2}\V,\Z,\bb)}{h}\right\}\m_\bb'(\W+\sqrt{-1}\bSigma^{1/2}\V,\Z,\bb)^{\otimes2}\right]\n\\
&&+
 E[
\psi\{Y-m(\W+\sqrt{-1}\bSigma^{1/2}\V,\Z,\bb)\}
\m_\bb''(\W+\sqrt{-1}\bSigma^{1/2}\V,\Z,\bb)
]\n\\
&=&\textstyle 
-2 E\left[
k_h\{Y-m(\X,\Z,\bb)\}\m_\bb'(\X,\Z,\bb)^{\otimes2}
\right]
 -E\left[\frac{Y-m(\X,\Z,\bb)}{h^2}k'\left\{\frac{Y-m(\X,\Z,\bb)}{h}\right\}\m_\bb'(\X,\Z,\bb)^{\otimes2}\right]\n\\
&&+E[\psi\{Y-m(\X,\Z,\bb)\}
\m_\bb''(\X,\Z,\bb)]\n\\
&=&\textstyle 
-2E \left[k_h\{Y-m(\X,\Z,\bb)\}
\m_\bb'(\X,\Z,\bb)^{\otimes2} 
\right] -E\left[\frac{Y-m(\X,\Z,\bb)}{h^2}k'\left\{\frac{Y-m(\X,\Z,\bb)}{h}\right\}\m_\bb'(\X,\Z,\bb)^{\otimes2}\right]\n\\
&& -C_2h^2E\{f_{\epsilon\mid\X,\Z}'(0,\X,\Z)\m_\bb''(\X,\Z,\bb)\}+O(h^4)\n\\
&=&  -2E\left( E
\left[k_h\left\{Y-m(\X,\Z,\bb)\right\}
\mid \X,\Z \right]\m_\bb'(\X,\Z,\bb)^{\otimes2} \right) \n\\
&& -E\left(E\left[\frac{Y-m(\X,\Z,\bb)}{h^2}k'\left\{\frac{Y-m(\X,\Z,\bb)}{h}\right\}\mid\X,\Z\right]\m_\bb'(\X,\Z,\bb)^{\otimes2}\right)
+ O(h^2)\n\\
&=& -2E( [\int_{-\infty}^\infty k(t) \{f_{\epsilon\mid\X,\Z}(0,
\X, \Z) + f'_{\epsilon\mid\X,\Z}(0, \X, \Z)ht\}dt +
O(h^2)]\m_\bb'(\X,\Z,\bb)^{\otimes2} )\n\\
&&-E([\int_{-\infty}^\infty tk'(t)\{f_{\epsilon\mid\X,\Z}(0, \X, \Z)
+f_{\epsilon\mid\X,\Z}'(0, \X, \Z)ht\}dt+O(h^2)]
    \m_\bb'(\X,\Z,\bb)^{\otimes2}
+ O(h^2)\n\\
&=&-
E\left\{f_{\epsilon\mid\X,\Z}(0,\X,\Z)
\m_\bb'(\X,\Z,\bb)^{\otimes2} + O(h^2)\right\} + O(h^2) \n\\
&=&-\A+O(h^2), 
\ee
and
\be \label{eq:acnorm}
&& E\left\{\|\frac{\partial  \S(Y, \W, \Z, \bb, \bSigma)}{\partial\bb\trans}\|_F^2\right\}\n\\
&=&E\left\{\| E(\frac{\partial[
\psi\{Y-m(\W+\sqrt{-1}\bSigma^{1/2}\V,\Z,\bb)\}
\m_\bb'(\W+\sqrt{-1}\bSigma^{1/2}\V,\Z,\bb)] }{\partial\bb\trans}
\mid
\W,\Z,Y)\|_F^2\right\}\n\\
&\le&E\left(\| \frac{\partial[
\psi\{Y-m(\W+\sqrt{-1}\bSigma^{1/2}\V,\Z,\bb)\}
\m_\bb'(\W+\sqrt{-1}\bSigma^{1/2}\V,\Z,\bb)]
}{\partial\bb\trans}\|_F^2\right)\n\\
&=&E\left\{E\left(\| \frac{\partial[
\psi\{Y-m(\W+\sqrt{-1}\bSigma^{1/2}\V,\Z,\bb)\}
\m_\bb'(\W+\sqrt{-1}\bSigma^{1/2}\V,\Z,\bb)]
}{\partial\bb\trans}\|_F^2\mid\X,\Z,Y\right)\right\}\n\\
&=&E\left(\| \frac{\partial[
\psi\{Y-m(\X,\Z,\bb)\}
\m_\bb'(\X,\Z,\bb)]
}{\partial\bb\trans}\|_F^2\right)\n\\
&\le&  12
E\left[\|
k_h\left\{Y-m(\X,\Z,\bb)\right\}
\m_\bb'(\X,\Z,\bb)^{\otimes2}\|_F^2
\right]\n\\
&& +3E\left[\|\frac{Y-m(\X,\Z,\bb)}{h^2}k'\left\{\frac{Y-m(\X,\Z,\bb)}{h}\right\}\m_\bb'(\X,\Z,\bb)^{\otimes2}\|_F^2\right]\n\\
&&+3
 E[\|
\psi\{Y-m(\X,\Z,\bb)\}
\m_\bb''(\X,\Z,\bb)\|_F^2
]\n\\
&=& h^{-1}C_3E\{f_{\epsilon\mid\X,\Z}(0,\X,\Z)
\|\m_\bb'(\X,\Z,\bb)^{\otimes2}\|_F^2\}+O(1),
\ee
where $C_3= 12\int k^2(t)dt +3\int t^2k'(t)^2dt$.
Thus
\bse 
&& \frac{1}{n}\sumi \frac{\partial \S(Y_i, \W_i, \Z_i, \bb, \bSigma)}{\partial\bb\trans}\Big|_{\bb=\bb^*}
=-\A+O_p\{h^2+(nh)^{-1/2}\},
\ese
and we have the simplification
 \be\label{eq:short}\textstyle 
&&n^{-1/2}\sumi  \S(Y_i, \W_i, \Z_i, \wh\bb_s, \bSigma)\n\\
&=&[-\A+O_p\{h^2+(nh)^{-1/2}\}]\sqrt{n}(\wh\bb_s-\bb)
+n^{-1/2}\sumi  \S(Y_i, \W_i, \Z_i, \bb, \bSigma).
\ee
Further,
\bse \label{eq:bc}
&&\var\{	 \S(Y_i, \W_i, \Z_i, \bb, \bSigma) \}\\
&=& E\left\{ \S(Y_i, \W_i, \Z_i, \bb, \bSigma)^{\otimes2}\right\}
-[E\{\S(Y_i, \W_i, \Z_i, \bb, \bSigma) \}]^{\otimes2}\\
&=&E\left[\left\{
E\left(\left[
    \tau-1+K\left\{\frac{Y_i-m(\W_i+\sqrt{-1}\bSigma^{1/2}\V,\Z_i,\bb)}{h}\right\}
\right.\right. \right.\right.
\n\\
&&\left.\left.\left.\left. 
+ \{Y_i-m(\W_i+\sqrt{-1}\bSigma^{1/2}\V,\Z_i,\bb) \} k_h\left\{Y_i-m(\W_i+\sqrt{-1}\bSigma^{1/2}\V,\Z_i,\bb)\right\} \right]\n\right.\right.\right.\\
&&\left.\left.\left.\times\m_\bb'(\W_i+\sqrt{-1}\bSigma^{1/2}\V,\Z_i,\bb) \mid
\W_i,\Z_i,Y_i\right)\right\}^{\otimes2}\right]+O(h^4)\\
&=&E\left[\left\{
E\left(\left[
\tau-1+I\left\{Y_i-m(\W_i+\sqrt{-1}\bSigma^{1/2}\V,\Z_i,\bb)\ge0\right\}\right]
\m_\bb'(\W_i+\sqrt{-1}\bSigma^{1/2}\V,\Z_i,\bb)\right.\right.\right.\\
&&\left.\left.\left. \mid
\W_i,\Z_i,Y_i\right)\right\}^{\otimes2}\right]+o(1)\\
&=&\B_{1}+o(1).
\ese
Combining  this with \eqref{eq:h2c}, under the Conditions \ref{con:bddregfunc}
and \ref{con:nh}, using central limit theorem, we obtain
\bse
\sqrt{n}(\wh\bb_s-\bb)
\sim N[-\A^{-1} C_2h^2n^{1/2}
E\{f_{\epsilon\mid\X,\Z}'(0,\X,\Z)
\m_\bb'(\X,\Z,\bb)\},
\A^{-1}\B_{1}\A\invT]
\ese
when $n\to\infty$. Note that in the above, $\|\B_{1} \|_2$ is bounded because
  \bse
  \|\B_{1}\|_2&\le&
\tr(\B_{1})\\
&=&  \tr E\left[\left\{
E\left(\left[
\tau-1+I\left\{Y_i-m(\W_i+\sqrt{-1}\bSigma^{1/2}\V,\Z_i,\bb)\ge0\right\}\right]
\m_\bb'(\W_i+\sqrt{-1}\bSigma^{1/2}\V,\Z_i,\bb)\right.\right.\right.\\
&&\left.\left.\left. \mid
      \W_i,\Z_i,Y_i\right)\right\}^{\otimes2}\right]\\
&\le&\tr E\left[
E\left\{ \left| \m_\bb'(\W_i+\sqrt{-1}\bSigma^{1/2}\V,\Z_i,\bb)\right|  \mid
  \W_i,\Z_i,Y_i\right\}^{\otimes2}\right]\\
&\le&\tr
E\left\{\left|\m_\bb'(\W_i+\sqrt{-1}\bSigma^{1/2}\V,\Z_i,\bb)\right|^{\otimes2}\right\}\\
&=&
E\left\{\|\m_\bb'(\W_i+\sqrt{-1}\bSigma^{1/2}\V,\Z_i,\bb)\|^2\right\}\\
&=&
 E\left[ E\left\{\|\m_\bb'(\W_i+\sqrt{-1}\bSigma^{1/2}\V,\Z_i,\bb)\|^2\mid\X_i,
  \Z_i, Y_i\right\}\right]\\
&=&
 E\left\{\|\m_\bb'(\X_i,\Z_i,\bb)\|^2\right\}<\infty,
 \ese
under Conditions \ref{con:bddregfunc}.  Further using  Condition \ref{con:nh}, we obtain
$\sqrt{n}(\wh\bb_s-\bb)\sim N(0,\A^{-1}\B \A\invT)$, when $n\to\infty$.
\qed

\subsection{Proof of Theorem \ref{th:classichat}} \label{apdx:th2pf}
We first show that $\wh\bb$ is a consistent estimator of $\bb$ by using Theorem 2.1 of \cite{newey1994large}.
Finding the solution of $n^{-1}\sumi \S(Y_i,\overline\W_i,\Z_i,	\bb, \wh \bSigma)\}=\0$
can be viewed as solving a maximization problem with objective function
	$\wh Q_n(\bb)\equiv -\| n^{-1}\sumi\S(Y_i, \overline\W_i,\Z_i, \bb,\wh \bSigma)\}
	\|_2^2$. Similar to the proof of Theorem \ref{th:classic}, let 
	$Q_0(\bb) \equiv -\| E\{\S(Y, \overline\W,\Z, \bb, \bSigma)\} \|_2^2$. Under
	Condition \ref{con:uniroot}, $Q_0(\bb)$ is uniquely maximized at the true
	$\bb$ in its neighborhood. Conditions \ref{con:compact} and
	\ref{con:bddregfunc} ensure that $\calB$ is compact and $Q_0(\bb)$
	is continuous on $\calB$. Lastly, $\wh Q_n(\bb)\to Q_0(\bb)$
	uniformly  in probability by the
	law of large numbers and the consistency of $\wh \bSigma$ in
        combination with the compactness of $\calB$.  
	Therefore, $\wh\bb\to\bb$ in probability when $n\to\infty$.

Next, a Taylor expansion leads to
\bse
\0&=&n^{-1/2} \sumi \S(\Y_i, \overline\W_i, \Z_i, \wh\bb, \wh\bSigma) \\
&=&n^{-1/2} \sumi \S(\Y_i, \overline\W_i, \Z_i, \wh\bb, \bSigma)
+\left\{n^{-1} \sumi \frac{\partial\S(\Y_i, \overline\W_i, \Z_i, \wh\bb,
    \bSigma^*)}{\partial\vech(\bSigma)\trans}\right\}n^{1/2}\vech(\wh\bSigma-\bSigma)\\
&=&n^{-1/2} \sumi \S(\Y_i, \overline\W_i, \Z_i, \bb, \bSigma)
+\{-\A+o_p(1)\}\sqrt{n}(\wh\bb-\bb)\\
&&+\left[E\left\{\frac{\partial\S(\Y_i, \overline\W_i, \Z_i, \bb,
      \bSigma)}{\partial \vech(\bSigma)\trans}\right\}+o_p(1)\right]
\frac{1}{\sqrt{n}}\sumi \vech (m^{-1}\M_i - \bSigma),
\ese
where $\bSigma^*$ is on the line connecting $\bSigma$ and
$\wh\bSigma$,
and
we used \eqref{eq:short}, the consistency of $\wh\bb$,
$\wh\bSigma$ and \eqref{eq:sigc} in the last equality above.
This leads to
\bse \textstyle 
\sqrt{n}(\wh\bb-\bb)
=\A^{-1} \frac{1}{\sqrt{n}}\sumi \left(\S(\Y_i, \ol\W_i, \Z_i, \bb, \bSigma)+E\left\{\frac{\partial\S(\Y_i, \overline\W_i, \Z_i, \bb,
      \bSigma)}{\partial \vech(\bSigma)\trans}\right\}\vech
  (m^{-1}\M_i - \bSigma)\right)+o_p(1),
\ese
where
\bse
&&\var \left(\S(\Y_i, \overline\W_i, \Z_i, \bb, \bSigma)+\left[E\left\{\frac{\partial\S(\Y_i, \overline\W_i, \Z_i, \bb,
      \bSigma)}{\partial \vech(\bSigma)\trans}\right\}\right]^{-1}\vech (m^{-1}\M_i - \bSigma)\right)\\
&=&E\{\S(\Y_i, \overline\W_i, \Z_i, \bb, \bSigma)^{\otimes2}\}\\
&&+E\left\{\frac{\partial\S(\Y_i, \overline\W_i, \Z_i, \bb,
      \bSigma)}{\partial \vech(\bSigma)\trans}\right\}E\left\{\vech (m^{-1}\M_i - \bSigma)^{\otimes2}\right\}E\left\{\frac{\partial\S(\Y_i, \overline\W_i, \Z_i, \bb,
      \bSigma)\trans}{\partial \vech(\bSigma)}\right\}\\
&=&\B_{1}+\B_{2}.
\ese
where the covariance term vanishes due to the independence between
$\M_i$ and $\Y_i, \overline\W_i, \Z_i$.
The central limit theorem thus leads to $\sqrt{n}(\wh\bb-\bb)
\to\MVN\{\0, \A^{-1}(\B_{1}+\B_{2})\A^{-\rm
	T}\}$ when $n\to\infty$.

\qed

 \subsection{Empirical bandwidth selection procedure}\label{sec:bandwidth}
 \begin{algorithm}[H]
 \caption{SIMEX based Bandwidth Selection} \label{algo1}
 \begin{algorithmic}[1]
 
 \Require Observed data $(Y_i, \W_i, \Z_i)_{i=1}^n$, measurement error variance $\bSigma$, the quantile regression function $m(\cdot)$, grid of bandwidths $\mathcal{H}$, set of inflation parameters $0 = \lambda_0 < \lambda_1 < \dots < \lambda_L$, number of simulations $B$
 
 \Ensure Selected bandwidth $\wh h$
 
 \For{$l = 1,\dots,L$}
     \For{$b = 1,\dots,B$}
     	\State Generate simulated measurement errors $\U_{ib} \sim \MVN(\0, \bSigma)$ for $i=1, ..., n$
		\State Construct pseudo-data: $\W_{ib}^{(l)} = \W_i + \sqrt{\lambda_l}\U_{ib}$
		\State Split data indexes into $K$ folds $\{\idxset_1, \dots, \idxset_K\}$.
     	\State Use $(K-1)$ folds of the data $(Y_i, \W_{ib}^{(l)},
        \Z_i)_{i=1}^n$ to estimate $\wh\bb_{-k}(h, \lambda_l)$ with measurement error level $(\lambda_l-\lambda_{l-1})\bSigma$, and then test on \textit{the held out fold} of the data $(Y_i, \W_{ib}^{(l-1)}, \Z_i)_{i=1}^n$ by quantile loss function. 
     	\State For each $h \in \mathcal{H}$, compute the  $K$-fold CV score:
     	                \bse
     	                CV_{\lambda_l}(h, b) =\frac{1}{K} \sum_{k=1}^K \frac{1}{|\idxset_k|}   \sum_{i \in \idxset_k} \rho_\tau[Y_i - m\{ \W_{ib}^{(l-1)}, \Z_i, \wh\bb_{-k}(h, \lambda_l)  \}].
     	                \ese
     \EndFor
     
     \State Compute average CV score: $\overline{CV}_{\lambda_l}(h) = B^{-1}\sum_{b=1}^{B} CV_{\lambda_l}(h, b)$
     
     \If{$\overline{CV}_{\lambda_l}(h)$ has an interior minimum at $h^*$}
         \State $\wh h(\lambda_l) = h^* \equiv \arg\min_{h} \overline{CV}_{\lambda_l}(h)$ \Comment{Standard U-shape selection}
     \Else
         \State Identify $h^*$ as the point of maximum curvature using \texttt{Kneedle} algorithm \cite{satopaa2011finding}
         \State $\wh h(\lambda_l) = h^*$ \Comment{Elbow-method for monotonic curves}
     \EndIf
 \EndFor
 
 \State Fit a robust linear model $g(\lambda; a, b) = a+ b\lambda$ to the points $\{\lambda_l, \log \wh h(\lambda_l)\}_{l=1}^L$
 \State Extrapolate to $\lambda_0$ to obtain the SIMEX bandwidth:  $\wh h = \exp\{g(\lambda_0; \wh a, \wh b)\}$
 \State \Return $\wh h$
 
 \end{algorithmic}
 \end{algorithm}
 
\subsection{Acknowledgments}
The authors would like to thank the anonymous referees, an Associate
Editor and the Editor for their constructive comments that improved the
quality of this paper.
The work is partially supported by National Institutes of Health, the Natural Sciences and Engineering Research Council of Canada, and Matthew Rosenshine Fund for Excellence in the Department of Statistics at Penn State.

\newpage
\bibliographystyle{plainnat}
\bibliography{quanref}

\end{document}